\documentclass[12pt,fleqn]{article} %[twocolumn]
\usepackage{graphicx}
\usepackage{textcomp} % for \textinterrobang
\usepackage{epstopdf}

\usepackage{latexsym}
\usepackage{amsmath}
\usepackage{bm}
\usepackage[mathcal]{euscript}
\usepackage{slashed}

\numberwithin{equation}{section}

\usepackage[top=30mm, bottom=40mm, left=25mm, right=25mm]{geometry}
\usepackage{marvosym}

\usepackage{indentfirst}
\newcommand\blfootnote[1]{
  \begingroup
  \renewcommand\thefootnote{}\footnote{#1}
  \addtocounter{footnote}{-1}
  \endgroup
}

\usepackage{hyperref}
\hypersetup{
        unicode,	% use with \texorpdfstring
        colorlinks,
        citecolor=blue,% filecolor=black,%
        linkcolor=blue, % urlcolor=black,% pdftex
        urlcolor=red,
        bookmarksopen=true,
        bookmarksopenlevel=\maxdimen,
}

\usepackage{amsmath, amssymb}    %Needed for align environment
\usepackage{bm}                  %For bold greek letter and such; command is \bm
\numberwithin{equation}{section}

\usepackage{amsthm} %Enable theorem environments
\theoremstyle{definition}

\theoremstyle{plain}

\newtheorem*{result*}{Result}

\def\gl#1#2{\ifmmode \mathrm{GL}(#1; {\bf #2}) \else $\mathrm{GL}(#1; {\bf #2})$\fi}
\def\sl#1#2{\ifmmode \mathrm{SL}(#1; {\bf #2}) \else $\mathrm{SL}(#1; {\bf #2})$\fi}
\def\so#1{\ifmmode \mathrm{SO}({#1}) \else $\mathrm{SO}(#1)$\fi}

\def\sp#1#2{\ifmmode \mathrm{Sp}(#1; {\bf #2}) \else $\mathrm{Sp}(#1; {\bf #2})$\fi}
\def\usp#1{\ifmmode \mathrm{USp}(#1) \else $\mathrm{USp}(#1)$\fi}
\def\spin#1{\ifmmode \mathrm{Spin}(#1) \else $\mathrm{Spin}(#1)$\fi}
\def\su#1{\ifmmode \mathrm{SU}({#1}) \else $\mathrm{SU}(#1)$\fi}
\def\double #1{#1{\hbox{\kern-2pt $#1$}}}
\mathcode`\*="702A                  % * now always complex conjugate
\def\half{{\textstyle{1\over{\raise.1ex\hbox{$\scriptstyle{2}$}}}}}

\def \p{\partial}

\def \e{\epsilon}

\def \St{{\rm Str}}

\def \psu{{\mathfrak{psu}}}

\def \frakh{{\mathfrak{h}}}
\def \frakg{{\mathfrak{g}}}
\def \frakm{{\mathfrak{m}}}
\def \Int{{\int\! {d}^2z\,}}

\begin{document}

%%%%% preprint number goes here
\begin{flushright}
\makebox[0pt][b]{}%{{\sc MI-TH-16xx}}
\end{flushright}

%%%%%%% title and authors
\vspace{40pt}
\begin{center}
{\LARGE Master symmetry in the $AdS_5\times S^5$ pure spinor string}

\vspace{30pt}
Osvaldo Chand\'ia,${}^{\clubsuit}$
William Divine Linch III,${}^{\heartsuit}$
and
Brenno Carlini Vallilo${}^{\spadesuit}$

\vspace{30pt}

{\em
${}^{\clubsuit}$ {\it Departamento de Ciencias, Facultad de Artes Liberales \&}\\
{\it Facultad de Ingenier\'{\i}a y Ciencias, Universidad Adolfo Ib\'a\~nez}\\
{\it Diagonal Las Torres 2640, Pe\~nalol\'en, Santiago, Chile}\\
\vspace{8pt}
${}^{\heartsuit}$ {\it George P. and Cynthia Woods Mitchell Institute for }\\
{\it Fundamental Physics and Astronomy, Texas A\&{}M University,}\\
{\it College Station, TX 77843-4242, USA}\\
\vspace{8pt}
${}^{ \spadesuit}$ {\it Departamento de Ciencias F\'{\i}sicas,
Universidad Andres Bello, \\Sazie 2212, Santiago, Chile}
}\\

\vspace{60pt}
{\bf Abstract}
\end{center}
%We show that the pure spinor description of the superstring in the
%$AdS_5\times S^5$ background has the set of classical non-local
%symmetries recently studied by Klose, Loebbert and M\"unkler in the
%context of $\mathbb Z_2$ cosets.
We lift the set of classical non-local
symmetries recently studied by Klose, Loebbert, and M\"unkler in the
context of $\mathbb Z_2$ cosets to
the pure spinor description of the superstring in the
$AdS_5\times S^5$ background.

%%%email addresses
\blfootnote{\\
${}^{\clubsuit}$  \href{mailto:ochandiaq@gmail.com}{ochandiaq@gmail.com}\\
${}^{\heartsuit}$ \href{mailto:wdlinch3@gmail.com}{wdlinch3@gmail.com}\\
${}^{\spadesuit}$ \href{mailto:vallilo@gmail.com}{vallilo@gmail.com} }

\setcounter{page}0
\thispagestyle{empty}

\newpage

\tableofcontents

\parskip = 0.1in
\section{Introduction}

Finding and studying integrable structures in the context of the AdS/CFT
correspondence has been one of the most active areas of research in
high energy physics. The theories on both sides of the conjecture
enjoy a large number of symmetries that make it possible to obtain
impressive results and checks of the conjecture. Although it lacks
more recent updates, a good review with an extensive list of references
is \cite{Beisert:2010jr}.  A more recent development not covered in
\cite{Beisert:2010jr} is the  the Quantum Spectral Curve method
\cite{Gromov:2013pga,Gromov:2014caa}. For some of its applications,
including higher loop computations, see \cite{Gromov:2014bva,
Alfimov:2014bwa,Marboe:2014gma,Marboe:2014sya,Gromov:2015wca,
Gromov:2015vua}.

In the famous work of Bena,
Polchinski, and Roiban \cite{Bena:2003wd}, it was shown that the
Green-Schwarz superstring in $AdS_5\times S^5$ \cite{Metsaev:1998it}
has an infinite set of classical conserved currents.
The existence of an analogous set of currents in the context of the
$AdS_5\times S^5$ pure spinor superstring was demonstrated in reference
\cite{Vallilo:2003nx}. Since this string is a generalization of
the usual $\mathbb Z_2$ coset to a super-coset with $\mathbb Z_4$
symmetry, the ability to lift this symmetry to the
super-coset is non-trivial.
In this note, we go one step further and show that
the pure spinor string in $AdS_5\times S^5$ admits an extension of the master
symmetry described by Klose, Loebbert, and M\"unkler
\cite{Klose:2016uur}. This symmetry complements the Yangian
symmetry,
%the lowest component of which acts as a
acting as a
raising operator on the classical Yangian charges.
The work presented here extends this structure to its super-analogue,
specifically, to the $\mathbb Z_4$ super-coset description
of the $AdS_5\times S^5$ pure spinor string.
In a sense the ghosts present in the pure
spinor string make the $\mathbb Z_4$ symmetry manifest with the ghosts' Lorentz
current playing the role of a gauge covariant current with vanishing
$\mathbb Z_4$ charge.

The classical and quantum integrability of the string in this
background has been explored much more for the GS string (see {\em
  e.g.} \cite{Arutyunov:2009ga}) than for the pure
spinor version. Some interesting results concerning the classical and
quantum integrability in the pure spinor formalism are given in references
\cite{Berkovits:2004jw,Berkovits:2004xu,Mikhailov:2007eg,Benichou:2011ch}.
A possible
application of
integrability techniques to the quantum pure spinor
string is to study its worldsheet dilatation operator
\cite{Ramirez:2015rma}. It has been shown that semi-classical
computations in the pure spinor string
give the same results as the GS string for a set of classical
solutions \cite{Aisaka:2012ud,Cagnazzo:2012uq}, but very little is known
about solutions dual to Wilson loops. This is an interesting line of
research to which the results presented in this work may have suitable
applications.

This paper is organized as follows: In section \ref{intro}, we give a
short review of the pure spinor string in $AdS_5\times S^5$ including
its flat current using a notation that will be useful in the subsequent
sections. In section \ref{master}, we extend the master symmetry
discussed in \cite{Klose:2016uur} to the pure spinor string.
In section \ref{yangian}, we derive how the existence of the
first Yangian charge
is a consequence of the master symmetry and the global $\psu(2,2|4)$
symmetry. We then give a general derivation of all higher
non-local and non-abelian charges the superstring has.
We conclude the paper and discuss directions for future research
in section \ref{conclu}.

\section{Pure spinor string in $AdS_5\times S^5$}
\label{intro}

The pure spinor string in the $AdS_5\times S^5$ background is described in
terms of the super-coset $PSU(2,2|4)/SO(1,4)\times SO(5)$. The Lie algebra
$\frakg=\mathfrak{psu}(2,2|4)$ is decomposed as $\frakg=
\bigoplus_{i=0}^3 \frakg_i$ with the projections satisfying
\begin{align}
  [\frakg_i,\frakg_j]\subset \frakg_{i+j\, {\rm mod }\, 4}.
\end{align}
The Killing form $\St(\cdot)$
also respects this symmetry in the sense that
%. This can be schematically written as
\begin{align}
   \St(\frakg_i\frakg_j) \neq 0\quad{\rm iff}\quad i+j = 0\;{\rm mod}\; 4.
\end{align}
This is a ${\mathbb Z}_4$ generalization of the usual ${\mathbb Z}_2$
symmetry present
in any symmetric space and, in particular, in the bosonic
coset construction.
For comparison and general convenience, we will define
\begin{align}
\frakh := \frakg_0
~~~\mathrm{and}~~~
\frakm := \bigoplus_{i=1}^3 \frakg_i .
\end{align}
Note that
$\frakh=\mathfrak{so}(1,4)\oplus\mathfrak{so}(5)$. An element $g$ of
the coset defines the left-invariant currents
\begin{align}\label{psucurrent}
  J = g^{-1}{d}g = K +A,
\end{align}
where ${d}=dz\p +d\bar z \bar\p$,\footnote{We will also use the
  notation $J=g^{-1}\p g$ and $\bar J=g^{-1}\bar\p g$. We hope the
  difference between the
  %two types of currents
  1-form current and its $dz$ component
  can be understood from
  context. }
$K\in \frakm$, and $A\in
\frakh$. We will also decompose $K= K_1+K_2+K_3$ with $K_i \in \frakg_i$ when
convenient. The gauge field $A$ is used in worldsheet covariant
derivatives $\nabla ={d} + [A,\,\cdot\, ]$. The Maurer-Cartan
identity
\begin{align}
  {d}J + J\wedge J=0
\end{align}
will also decompose into four independent identities along each
$\frakg_i$.

In addition to the geometric part, the
pure spinor string is defined with pure spinor ghosts and their conjugate momenta.
%The ghosts and their conjugate momenta
These are invariant under global $PSU(2,2|4)$
transformations.
The ghosts are
fermionic elements of the algebra
\begin{align}
  \lambda \in \frakg_1
  \quad \mathrm{and} \quad
  \bar\lambda \in \frakg_3
\end{align}
that satisfy
\begin{align}
  \{\lambda,\lambda\} = 0= \{\bar\lambda,\bar\lambda\}.
\end{align}
This is the coset generalization of the pure spinor condition in flat space.
%which are equivalent to the pure spinor condition in flat space.
The momenta conjugate to the pure spinor variables are denoted
\begin{align}
  \omega \in \frakg_3
  \quad \mathrm{and} \quad
  \bar\omega \in \frakg_1.
\end{align}
They suffer the gauge transformations
\begin{align}
  \delta\omega = [A, \lambda]
  \quad \mathrm{and} \quad
  \delta\bar\omega = [B,\bar\lambda],
\end{align}
where $A$ and $B$ are any two local bosonic elements of $\frakg_2$. We will
also define
\begin{align}
  N=-\{\lambda,\omega\}
  \quad \mathrm{and} \quad
  \bar N=-\{\bar\lambda,\bar\omega\}
\end{align}
which are the Lorentz generators for the ghosts. Note that they have
zero ${\mathbb Z}_4$ charge. The pure spinor
condition implies
\begin{align}
  [\lambda,N]=0=[\bar\lambda,\bar N].
\end{align}

Having all the ingredients, we can write the pure spinor action
\cite{Berkovits:2000fe,Berkovits:2000yr,Vallilo:2002mh}
\begin{align}\label{action}
  S=\frac14 \int\!\! d^2z\, \St\left( K_1\bar K_3+2 K_2\bar K_2
  +3 K_3\bar K_1-4 N\bar N +4 \omega \bar\nabla \lambda +
 4 \bar\omega\nabla\bar\lambda  \right).
\end{align}
The geometric part of this action is the standard kinetic term of a
coset model plus a Wess-Zumino term
\begin{align}
  S_{\rm WZ}= -\frac14 \Int \St\left( K_1\bar K_3 - K_3\bar K_1\right).
\end{align}
This particular coefficient of the Wess-Zumino term is fundamental for BRST
symmetry and integrability \cite{Vallilo:2003nx,Berkovits:2004xu}.

By construction, the action has global $PSU(2,2|4)$ invariance and local
$SO(1,4)\times SO(5)$ invariance. Global transformations act on $g$ by
left multiplication and the local transformations act on $g$ by right
multiplication. The current $J$ is invariant under the
global symmetry. On the other hand, $K$ tranforms in the adjoint
representation of $\frakh$ if $\delta g= gM$, where $M\in\frakh$
and  $A$ transforms as a connection. The ghosts and their conjugate
momenta transform in the adjoint representation of $\frakh$ as well.

The next fundamental symmetry is BRST invariance
defined by\footnote{These transformations are nilpotent only up to local
$SO(1,4)\times SO(5)$ transformations and equations of motion. There are ways to
fix both these issues
\cite{Berkovits:2007rj,Bedoya:2010qz,Chandia:2006ix,Chandia:2014sta},
however, they will not be needed here.
} %end footnote
\begin{align}
  \delta g = g (\lambda +\bar\lambda),\quad
  \delta \lambda =0,\quad
  \delta \bar\lambda=0, \quad
  \delta \omega =- K_3,\quad \delta\bar\omega =- \bar K_1.
\end{align}
The conserved current
associated with BRST symmetry is given by
\begin{align}
  j_{\rm BRST}= \St(\lambda K_3) dz + \St(\bar\lambda\bar K_1)d\bar
  z.
\end{align}
It is not only conserved ${d}\ast j_{\rm BRST}=0$, but its components are holomorphic and
anti-holomorphic
\begin{align}
  \bar\p\big( \St(\lambda K_3)\big)=0
  \quad \mathrm{and}\quad
  \p\big( \St(\bar\lambda\bar K_1)\big)=0
\end{align}
after using the equations of motion which will be discussed below.
This fact  means that the charges defined by
\begin{align}
  Q_\e = \oint\! dz\, \e(z)\St(\lambda K_3)
  \quad \mathrm{and}\quad
  Q_{\bar\e} = \oint\! d\bar
  z\, \bar\e(\bar z) \St(\bar\lambda\bar K_1)
\end{align}
also generate symmetries for any two {independent} holomorphic and anti-holomorphic
functions $\e(z)$ and $\bar\e(\bar z)$. In this case the BRST
transformations above generalize to\footnote{
It may seem surprising that the BRST invariance in the pure spinor
superstring implies a much larger symmetry than the usual BRST symmetry in field theory.
However, we should remember that the pure spinor BRST should also imply
Virasoro symmetry which is an infinite-dimensional chiral symmetry.
} %end footnote
\begin{align}
%  \delta g &= g \big[\e(z)\lambda +\bar\e(\bar z)\bar\lambda\big],\quad
%  \delta \lambda =0,\quad
%  \delta \bar\lambda=0,\nonumber \\
%  \delta \omega& =- \e(z) K_3,\quad \delta\bar\omega =-\bar\e(\bar z) \bar K_1.
  \delta g &= g \big[\e(z)\lambda +\bar\e(\bar z)\bar\lambda\big],\quad
  \delta \lambda =0,\quad
  \delta \bar\lambda=0,\quad
  \delta \omega& =- \e(z) K_3,\quad \delta\bar\omega =-\bar\e(\bar z) \bar K_1.
\end{align}

We now compute the current associated with the global
$PSU(2,2|4)$ symmetry. The coset element transforms as $\delta g
=\Omega g$. We will let $\Omega$ be a local parameter and use the
Noether method. The left invariant currents transform as
\begin{align}
  \delta K_i = g^{-1}({d}\Omega) g\Big|_{\frakg_i},\quad \delta A =
  g^{-1}({d}\Omega) g \Big|_{\frakh}.
\end{align}
When inserting this transformation into the action, we can drop the
restriction on the subspaces since the transformations will always
come together with a dual algebra element inside a supertrace. The
action transforms as
\begin{align}
~\hspace{-2mm} \delta S=\frac14 \int\! d^2z\, \St\left(  g^{-1}\p\Omega g \bar
  K_3+  g^{-1}\bar\p\Omega g K_1 +\cdots +4 g^{-1}\p\Omega g \bar
  N+ 4g^{-1}\bar\p\Omega g N\right),
\end{align}
from which we read off the Noether current
\begin{align}\label{globalC}
~\hspace{-1mm}  {\rm j} =g \left( K_1 + 2K_2 + 3 K_3 +4N \right)g^{-1}
  dz + g\left(3 \bar K_1+2\bar K_2 + \bar K_3 +  4\bar N \right)g^{-1} d\bar z.
\end{align}
Conservation of the current ${d}\ast {\rm j}=0$ implies the equations of
motion which can be written compactly as
\begin{align}\label{EOMs}
  [ \bar\nabla +\bar K , K_1 + 2K_2 + 3 K_3 +4N]+[\nabla
  +K, 3 \bar K_1+ 2\bar K_2 + \bar K_3 + 4\bar N]=0.
\end{align}
These equations are calculated by varying the action (\ref{action}) with
respect to a variation of the coset element given by $\delta g
= gX$ with $X\in \frakm$.
Using the ${\mathbb Z}_4$ decomposition, the Maurer-Cartan identity
for $J$, and
\begin{align}
  \nabla \bar N - \bar\nabla N -2 [N,\bar N]=0,
\end{align}
we can separate (\ref{EOMs}) into eight equations of motion. To derive
this last equation we use the equations of motion for the ghosts
coming from (\ref{action}):
\begin{align}
  [\bar\nabla -\bar N ,\lambda +\omega]=[\nabla - N,\bar\lambda +\bar\omega]=0.
\end{align}

We can use an operator $\Sigma$  defined in \cite{Beisert:2012ue}\ to
write very compact expressions for the action and other observables.
We define the action of $\Sigma$ on the basic currents as
\begin{align}
  \Sigma(A)=0,\quad \Sigma(K_1)=K_1,\quad \Sigma(K_2)=2K_2,\quad
  \Sigma(K_3)=3K_3,\quad \Sigma(N)=4N,\\
  \Sigma(\bar A)=0,\quad \Sigma(\bar K_1)=3\bar K_1,\quad
  \Sigma(\bar K_2)=2\bar K_2,\quad
  \Sigma(\bar K_3)=\bar K_3,\quad \Sigma(\bar N)=4\bar N.
\end{align}
Then, the action can be written as
\begin{align}
  S=\int\!\!d^2z \Big(\frac14 \bar K\,\Sigma( K)  +
  \omega\bar\nabla\lambda+\bar\omega\nabla\bar\lambda-N\bar N \Big),
\end{align}
and the components of the Noether current can be written as
\begin{align}
  j_z= g \Sigma(K+N) g^{-1},\quad j_{\bar z}= g\Sigma(\bar
  K+\bar N)g^{-1}.
\end{align}
For the supertrace, we have
\begin{align}
  \St\big( O_i \Sigma(\bar O_j)\big)=\St\big(\Sigma(O_i)\bar O_j\big),
\end{align}
where $O_i$
is any current with a defined action of $\Sigma$.
We note, however, that the usefulness of $\Sigma$ in computations is
limited by the fact that it is not a Lie algebra homomorphism ({\it e.\ g.} it does not
preserve the Lie bracket).

\subsection{The flat current}
In contrast to the Noether current of the bosonic cosets,
the conserved current (\ref{globalC}) of the $\mathbb Z_4$ super-coset is not flat.
Instead, it was shown in reference \cite{Vallilo:2003nx}
%by one of the authors
that the pure spinor string in $AdS_5\times S^5$ has a family of flat
currents
depending on a complex parameter $\mu$:
\begin{align}\label{superflat}
  {\mathcal L}_\mu = l_\mu dz + \bar l_\mu d\bar z ,
\end{align}
with
\begin{align}
l_\mu &=
        g\left[
                (e^{2\mu}-1) K_2 +
                (e^{\mu} -1 ) K_1 +
                (e^{3\mu} -1 ) K_3 +
                (e^{4\mu}-1) N
        \right]g^{-1}
\cr
\bar l_{\mu} &=
        g\left[
                (e^{-2\mu}-1) \bar K_2 +
                (e^{-3\mu} -1 ) \bar K_1 +
                (e^{-\mu} -1 ) \bar K_3 +
                (e^{-4\mu}-1) \bar N
        \right]g^{-1} .
\end{align}
Using the $\Sigma$ operator defined above, we can write this compactly as
\begin{align}
  l_\mu = g\left[ e^{\mu\Sigma}-1\right](J+N)g^{-1}
  \quad \mathrm{and}\quad
  l_\mu = g\left[ e^{-\mu\Sigma}-1\right](\bar J+\bar N)g^{-1} .
\end{align}
The current is flat
\begin{align}\label{flatcurrent}
  {d} {\mathcal L}_\mu + {\mathcal L}_\mu \wedge {\mathcal L}_\mu =0
\end{align}
as a consequence of the equations of motion.

The existence of this current is remarkable given that,
as just mentioned, the conserved current of a ${\mathbb Z}_4$
super-coset is generally not flat. In particular, there is no
value of $\mu$ for which the flat current (\ref{superflat})
reduces to (\ref{globalC}). However, note that
\begin{align}\label{niceEq}
   %{\mathcal L}'_\mu\Big|_{\mu=0}
   {\mathcal L}'_{0}
   =\ast{\rm j},
\end{align}
where $'=\frac{{d}}{{d}\mu}$ and the left-hand side is evaluated
at $\mu=0$. This is the usual statement that the first charge generated by the
flat current ${\mathcal L}_\mu$ is the conserved charge of the
$\psu(2,2|4)$ algebra. This can be seen from the monodromy
matrix\footnote{To avoid global issues, we will assume the
  worldsheet is infinite and open.}
\begin{align}\label{monodromy}
  \bm{M}(\mu)= {\rm P}\,{\rm exp}\Big( \int_{-\infty}^\infty {\mathcal L}_\mu\Big).
\end{align}
The charge of the $\psu(2,2|4)$ algebra
\begin{align}
    \bm{Q}_{\psu}= \int_{-\infty}^\infty \ast {\rm j}
\end{align}
is the coefficient of the first power of $\mu$ in the expansion of $\bm{M}(\mu)$.
% However, we can use the flat current to show that the commutator of its
% components is a total derivative
% \begin{align}
%   [ j_z , j_{\bar z} ] = -[ l'_0, \bar l'_0]
%   = \p\bar l''_0  - \bar\p l''_0 .
% \end{align}
% Here we used the fact that $l_0=\bar l_0=0$.

\section{Master symmetry}
\label{master}

Following Klose, Loebbert, and M\"unkler \cite{Klose:2016uur}, we can
define a flat deformation of the Maurer-Cartan current by
\begin{align}
  L_\mu = J + g^{-1}{\mathcal L}_\mu g =\left[A+
                e^\mu K_1 +
                e^{2\mu} K_2 +
                e^{3\mu} K_3 +
                (e^{4\mu}-1) N
        \right]dz \cr
        +\left[ \bar A +
                e^{-3\mu} \bar K_1 +\
                e^{-2\mu} \bar K_2 +
                e^{-\mu} \bar K_3 +
                (e^{-4\mu}-1) \bar N
        \right]d\bar z .
\end{align}
Note that $L_0 = J$, since ${\mathcal L}_0=0$.
Using the $\Sigma$ operator defined in the previous section,
$L_\mu$ can be written as
\begin{align}
  L_\mu=\left[ e^{\mu\Sigma}(J+N) -N\right]dz+ \left[
  e^{-\mu\Sigma}(\bar J+\bar N) -\bar N\right]d\bar z.
\end{align}
(It is actually more straightforward to verify that $L_\mu$ satifies a flatness
condition.)

A deformation $g_\mu$ of the coset element $g$ can be defined by the
differential equation \cite{Eichenherr:1979ci,Brezin:1979am}
\begin{align}\label{defg1}
  {d} g_\mu(z,\bar z)= g_\mu(z,\bar z)L_\mu\quad {\rm with }\quad
  g_\mu(z_0,\bar z_0)=g(z_0,\bar z_0).
\end{align}
Here, $(z_0,\bar z_0)$ is any reference point on the
worldsheet needed to fix an ``initial
condition''. This equation is well-defined since $L_\mu$ is flat.
Consequently, this deformation of $g$ is only defined on-shell.  An
ansatz to solve it is
\begin{align}\label{defg2}
  g_\mu(z,\bar z) =\chi_\mu(z,\bar z) g(z,\bar z),
\end{align}
where $\chi_\mu$ satisfies
\begin{align}\label{chidef}
  {d}\chi_\mu =  \chi_\mu{\mathcal L}_\mu \quad{\rm with}\quad
  \chi_\mu(z_0,\bar z_0)=\bm{1}.
\end{align}
Again, this equation is well defined since ${\mathcal L}_\mu$ is
flat.
We can expand this differential equation in a power series about $0$.
The flat current vanishes for $\mu=0$, so the first
two equations are
\begin{align}\label{chiexp}
  {d}\chi^{(0)}=0,\quad {d}\chi^{(1)}= \ast{\rm j}
  = j_z dz - j_{\bar z} d\bar z,
\end{align}
where we used (\ref{niceEq}).
Written in
this way, it is clear that $\chi^{(1)}$ only exists if the
equations of motion are satisfied. The solution is given by
\begin{align}
 \chi^{(0)}(z,\bar z)=\bm{1},\quad  \chi^{(1)}(z,\bar z) = \int_{(z_0,\bar z_0)}^{(z,\bar z)}(dz j_z-d\bar z j_{\bar z}).
\end{align}
With this, we are finally in the position to define the
``master symmetry'' \cite{Klose:2016uur}
\begin{align}
  \hat\delta g(z,\bar z) := \chi^{(1)}(z,\bar z)g(z,\bar z).
\end{align}
This is a non-local transformation acting on the currents as
\begin{align}\label{nonLocalT}
  \hat\delta J=g^{-1}{d}\chi^{(1)}g = \hat\delta K_2+ \hat\delta K_1+
  \hat\delta K_3+ \hat\delta A = g^{-1}( j_zdz -j_{\bar z}d\bar
  z)g=\nonumber  \\
  \left( K_1 + 2K_2 + 3 K_3 + 4 N\right)
  dz - \left(\bar K_3+2\bar K_2 +3 \bar K_1 +4 \bar
  N\right) d\bar z
\end{align}

Up until this point, the master symmetry has been discussed at the level of
geometry. The pure spinor string also has ghosts, and we should consider
that $\hat\delta$ also acts on them. Since the structure under discussion is
on-shell, we will consider $N$ and $\bar N$ as fundamental fields with
defining equations of motion $\nabla\bar N + \bar\nabla N=0$ and
$\nabla\bar N - \bar\nabla N -2 [N,\bar N] =0$. We will define the
extension of the master symmetry to act on them as
\begin{align}\label{masterGhost}
  \hat\delta N = 4N, \quad \hat\delta \bar N = -4\bar N.
\end{align}

With these transformations, it is immediate to
verify that (\ref{nonLocalT}) is a symmetry of the
equations of motion (\ref{EOMs}), turning it into a Maurer-Cartan
identity together with $\nabla \bar N - \bar\nabla N -2 [N,\bar
N]=0$. As argued in \cite{Eichenherr:1979ci},
$\chi_\mu$ generates an infinite tower of non-local
symmetries of a ${\mathbb Z}_2$ coset. The analogous statement for
the pure spinor string for any value of
$\mu$ will be proved at the end of this paper. The main difference
here, apart from the presence of fermionic terms, is that the
gauge field $A$ transforms into the ghost current $N$, thereby
mixing matter and ghosts.

Since this symmetry is only defined on-shell, discussing it at the
level of the action is potentially meaningless. Nevertheless,
we follow reference \cite{Klose:2016uur}\ and
try to use Noether procedure to calculate the current
associated to the master symmetry anyway. We include a local parameter
$\epsilon(z,\bar z)$ in the transformation of $g$
\begin{align}
  \hat\delta g = \epsilon(z,\bar z)\chi^{(1)}g.
\end{align}
The currents transform as
\begin{align}
  \hat\delta J = {d}\epsilon g^{-1}\chi^{(1)}g +
  \epsilon\Sigma \left( K+ N\right)
  dz -\epsilon\Sigma \left(\bar K +\bar N\right) d\bar z.
\end{align}
Inserting this transformation into the action and only collecting
terms depending on ${d}\epsilon$, we find
\begin{align}
  \hat\delta S =\frac14 \Int\St\Big(\p\epsilon\chi^{(1)}j_{\bar z} +
  \bar\p\epsilon\chi^{(1)}j_z\Big).
\end{align}
Using the same normalization as (\ref{globalC}),
we read off the conserved current
\begin{align}\label{nonlocalC}
  \bm{J}^{(0)} = \St\Big(\chi^{(1)}j_z\Big)dz +
  \St\Big(\chi^{(1)}j_{\bar z}\Big)d\bar z.
\end{align}
Note that by (\ref{chiexp}),
\begin{align}
 \ast\bm{J}^{(0)} = \frac12 {d}\big(\St(\chi^{(1)}\chi^{(1)})\big).
\end{align}
Using this, we can perform the integral to find the conserved
charge
\begin{align}
  {\mathfrak C}^{(0)} = \int \ast\bm{J}^{(0)}=
  \frac12 \St(\chi^{(1)}\chi^{(1)})\Big|_{-\infty}^\infty.
\end{align}
If we choose the point $z_0$ in the initial condition (\ref{chidef}) of
$\chi_\mu(z,\bar z)$ to be at
spacial $-\infty$ and use that the $\psu(2,2|4)$ charge is
given by
\begin{align}
  \bm{Q}_{\psu}= \chi^{(1)}(\infty),
\end{align}
we see that
\begin{align}
  {\mathfrak C}^{(0)} = \frac12 \St\big(\bm{Q}_{\psu}\bm{Q}_{\psu})
\end{align}
is the Casimir of the $\psu(2,2|4)$ algebra,
as in \cite{Klose:2016uur}.

In principle we could find the higher scalar charges associated
with higher powers of the $\mu$ expansion of $\chi_\mu$ in a similar way,
but
%it is faster to guess the result.
we can guess the result as follows.
The form in (\ref{nonlocalC}) suggests that the scalar current
containing all higher master symmetry charges is
$\ast\St\big( \chi_\mu {\mathcal L}'_\mu\big)$. However, there are a
few problems with this first attempt: $\chi_\mu$ is not
an element of the algebra, it is not conserved, and its
$\mu^2$ coefficient does not match the Casimir we obtain from acting
with $\hat\delta$ on ${\mathfrak C}^{(0)}$. A better guess is
\begin{align}
  \bm{J}_\mu = \ast\St\big( \chi^{-1}_\mu\chi_\mu' {\mathcal L}'_\mu\big).
\end{align}
This is conserved $d\ast \bm{J}_\mu=0$ using that
$d(\chi^{-1}_\mu\chi_\mu')=[\chi^{-1}_\mu\chi_\mu', {\mathcal
  L}_\mu]+{\mathcal L}_\mu'$ and the first derivative with
respect to $\mu$ of the flatness condition (\ref{flatcurrent}).
Then, the complete tower of nonlocal charges
can be defined by
\begin{align}\label{nonlocalCasimir}
  {\mathfrak C}_\mu=
  \int_{-\infty}^\infty \ast \bm{J}_\mu
  =\int_{-\infty}^\infty\St\big( \chi^{-1}_\mu\chi_\mu' {\mathcal
  L}'_\mu\big)
  =\frac12 \St\Big(
  (\chi^{-1}_\mu\chi_\mu'(\infty))(\chi^{-1}_\mu\chi_\mu'(\infty))\Big).
\end{align}
The zeroth power of $\mu$ in the expansion gives the Casimir
${\mathfrak C}^{(0)}$, and it is easy to show that the coefficient of
$\mu$ is the result of calculating $\hat\delta {\mathfrak C}^{(0)}$.

\section{Yangians}
\label{yangian}
Having defined $\hat\delta$, we can see how it affects other symmetries
of the string. If we act with $\hat\delta$ on (\ref{globalC}) we obtain
\begin{align}\label{nonLocal0}
  \hat\delta {\rm j}={\rm j}^{(1)} &= g (K_1+4K_2+9K_3+16N)g^{-1} {d}z
        \cr
        &-g(9\bar K_1+4\bar K_2+\bar K_3 +16\bar N)g^{-1} {d}\bar z
         + [ \chi^{(1)} , {\rm j} ],
\end{align}
which is the first non-local current given by the
monodromy matrix (\ref{monodromy})
constructed from the flat current (\ref{superflat}).
Schematically, %It seems that
this current generates a transformation on the coset element of the
form $\delta g \sim [\eta, \chi^{(1)}]g$,
where $\eta\in\psu(2,2|4)$ is constant.
We can try to use the Noether method again to see if we can obtain
the non-local current as the Noether current associated with this
transformation. However,
% if we go through that we will only obtain the
carrying this out, we obtain only the
last term in (\ref{nonLocal0}).
As mentioned previously, it is not surprising that this time we
could not obtain the desired result since this is an on-shell
symmetry. In the case of the principal chiral model it is possible to
interpret these non-local currents as Noether currents
\cite{Dolan:1980kz,Hou:1981hn}, but it is not clear that we can use
the same method here.

We could obtain the higher non-local currents by
successive applications of the master symmetry generator $\hat\delta$,
but as before there is a faster way to obtain these currents as we now show.

\subsection{Non-local current}

The non-local current associated with the global symmetry and all
higher Yangian charges is calculated by replacing
$g\to g_\mu$ (\ref{defg1}, \ref{defg2}) in the definition
of the Noether current (\ref{globalC}) and defining a finite $\mu$
deformation of ghost currents as
\begin{align}
  N\to N_\mu=e^{4\mu}N,\quad \bar N\to\bar N_\mu =e^{-4\mu}\bar N,
\end{align}
from which we can see that the master symmetry (\ref{masterGhost})
corresponds to the first power of the
$\mu$ deformation. With these definitions we have
\begin{align}
  {\mathbb J}_\mu &= g_\mu( (K_\mu)_1 + 2(K_\mu)_2 + 3(K_\mu)_3 +
                     4N_\mu)g^{-1}_\mu{d}z\nonumber \\
                  &+g_\mu( (\bar K_\mu)_1 + 2(\bar K_\mu)_2 + 3(\bar K_\mu)_3 +
   4\bar N_\mu)g^{-1}_\mu{d}\bar z
                   \cr &=
                    \chi_\mu g (e^{\mu}K_1+2e^{2\mu}K_2+3e^{3\mu}K_3+
  4e^{4\mu}N)g^{-1}\chi^{-1}_u {d}z\nonumber \\
  &+\chi_\mu g (3e^{-3\mu}\bar K_1+2e^{-2\mu}\bar K_2+e^{-\mu}\bar K_3+
  4e^{-4\mu}\bar N)g^{-1}\chi^{-1}_u {d}\bar z,
\end{align}
where we defined $K_\mu=(g^{-1}_\mu{d}g_\mu)|_\frakm =
(g^{-1}{\mathcal L}_\mu g + g^{-1}{d}g)|_\frakm $. From the last two
lines we can identify this current as
\begin{align}
  {\mathbb J}_\mu= \chi_\mu\big(\ast {\mathcal L}'_\mu\big) \chi^{-1}_\mu.
\end{align}
Computing the $\p$ and $\bar\p$ derivatives of its components, we see that
\begin{align}
  \bar\p {\mathbb J}_\mu &= \chi_\mu\Big(g[ \bar\nabla +\bar K ,
                            2e^{2\mu}K_2+e^{\mu}K_1+3e^{3\mu}K_3+
       4e^{4\mu}N]g^{-1}
\cr
&+ [\bar l_\mu , g( 2e^{2\mu}K_2+e^{\mu}K_1+3e^{3\mu}K_3+
       4e^{4\mu}N)g^{-1}]\Big)\chi^{-1}_\mu
\cr
  &= \chi_\mu\Big(\bar\p l'_\mu  +
     \big[ \bar l_\mu ,l'_\mu\big]\Big)\chi^{-1}_\mu.
\end{align}
\begin{align}
  \p \bar {\mathbb J}_\mu &= \chi_\mu\Big(
                             g[ \nabla + K ,  2e^{-2\mu}\bar K_2+3e^{-3\mu}\bar
  K_1+e^{-\mu}\bar K_3+ 4e^{-4\mu}\bar N]g^{-1}
  \cr
  &+ [ l_\mu , g( 2e^{-2\mu}\bar
  K_2+3e^{-3\mu}\bar K_1+e^{-\mu}\bar K_3+
  4e^{-4\mu}\bar N)g^{-1}]\Big)\chi^{-1}_\mu
\cr
&=\chi_\mu\Big( -\p \bar l'_\mu  +
     \big[  l_\mu ,-\bar l'_\mu\big] \Big)\chi^{-1}_\mu.
\end{align}
So, the conservation of ${\mathbb J}_\mu$ is simply the first
derivative with respect to $\mu$
of the flatness condition (\ref{flatcurrent}) of ${\mathcal L}_\mu$:
\begin{align}
  \bar\p {\mathbb J}_\mu + \p\bar{\mathbb J}_\mu =
  \chi_\mu\Big(
  \bar\p l'_\mu  -\p \bar l'_\mu + \big[ \bar l_\mu ,l'_\mu\big] +
  \big[\bar l'_\mu,l_\mu \big] \Big)\chi^{-1}_\mu = 0.
\end{align}
This relation proves that if $g$ is a solution, then the deformation $g_\mu$ is also a solution.
%is a symmetry of the original on-shell $g$.
The current ${\mathbb J}_\mu$ contains a whole
tower of non-local conserved currents of the model, starting with the global
$\psu(2,2|4)$ current (\ref{psucurrent}). It is easily checked that
\begin{align}
  {\mathbb J}_\mu= {\rm j} + \mu {\rm j}^{(1)} +\cdots ,
\end{align}
where ${\rm j}^{(1)}$ is the first Yangian current. To prove
the higher $\mu$ powers are all higher Yangian currents and that
$\hat\delta$ acts as a raising operator, we proceed as
follows. First we note that
\begin{align}
  \hat\delta {\mathcal L}'_\mu ={\mathcal L}''_\mu +
  [\chi^{(1)},{\mathcal L}'_\mu].
\end{align}
After some manipulations, one can show that
\begin{align}
%  \hat\delta {\mathbb J}_\mu =
%  \chi_\mu\big( \ast{\mathcal L}''_\mu\big)\chi^{-1}_\mu
%  + \chi_\mu [ \chi^{-1}_\mu\hat\delta\chi_\mu ,
%  \ast {\mathcal L}'_\mu]\chi^{-1}_\mu
%  +\chi_\mu[\chi^{(1)},{\mathcal L}'_\mu]\chi^{-1}_\mu .
  \chi^{-1}_\mu \hat\delta {\mathbb J}_\mu  \chi_\mu =
  \ast{\mathcal L}''_\mu
  + [ \chi^{-1}_\mu\hat\delta\chi_\mu ,
  \ast {\mathcal L}'_\mu]
  +[\chi^{(1)},{\mathcal L}'_\mu].
\end{align}
To get to final result let us now compute the derivative with respect
to $\mu$ of ${\mathbb J}_\mu$
\begin{align}
%  {\mathbb J}'_\mu = \chi_\mu\big( \ast{\mathcal L}''_\mu\big)\chi^{-1}_\mu
%  + \chi_\mu [ \chi^{-1}_\mu\chi'_\mu , \ast {\mathcal L}'_\mu]\chi^{-1}_\mu.
  \chi^{-1}_\mu {\mathbb J}'_\mu \chi_\mu=  \ast{\mathcal L}''_\mu
  + [ \chi^{-1}_\mu\chi'_\mu , \ast {\mathcal L}'_\mu] .
\end{align}
If we subtract both equations we have
\begin{align}
  \chi^{-1}_\mu \big( \hat\delta {\mathbb J}_\mu- {\mathbb J}'_\mu \big) \chi_\mu=
   [ \chi^{-1}_\mu\hat\delta\chi_\mu
  -\chi^{-1}_\mu\chi'_\mu, \ast {\mathcal L}'_\mu]
  +[\chi^{(1)},\ast {\mathcal L}'_\mu].
\end{align}

Let us call $\phi_\mu :=\chi^{-1}_\mu\hat\delta\chi_\mu-\chi^{-1}_\mu\chi'_\mu$ and note that
$\phi_0= -\chi^{(1)}$. Using that
\begin{align}
  \hat\delta {\mathcal L}_\mu ={\mathcal L}'_\mu +[\chi^{(1)},{\mathcal
  L}_\mu]-\ast{\rm j},
\end{align}
we can calculate that the differential equation satisfied by
$\phi_\mu$ is
\begin{align}
  { d}\phi_\mu +\ast {\rm j} = [\phi_\mu+ \chi^{(1)} ,{\mathcal
  L}_\mu].
\end{align}
Since $\ast{\rm j}={ d}\chi^{(1)}$, we can change variables $\phi_\mu \to \psi_\mu:=\phi_\mu+ \chi^{(1)}$
and
arrive at
\begin{align}
  { d}\psi_\mu = [\psi_\mu , {\mathcal L}_\mu],
\end{align}
where now the initial condition is $\psi_0=0$. Since ${\mathcal
  L}_0=0$, all higher powers of $\mu$ will vanish. So we conclude that
$\phi_\mu = -\chi^{(1)}$ for any value of  $\mu$.\footnote{This result can be used as a practical way to
  calculate $\hat\delta \chi_\mu= \chi_\mu' -\chi_\mu \chi^{(1)}$ for
  any power of $\mu$.}
Thus, we have finally proven that
\begin{align}
  \hat\delta {\mathbb J}_\mu =\frac{ d}{d\mu}{\mathbb
  J}_\mu,
\end{align}
so that $\hat\delta$ acts as a raising operator for non-local
symmetires, exactly as in \cite{Klose:2016uur}. Similarly, it can be
shown that the non-local Casimir (\ref{nonlocalCasimir}) satisfies
\begin{align}
  \hat\delta {\mathfrak C}_\mu = \frac{ d}{d\mu}{\mathfrak C}_\mu.
\end{align}

\section{Conclusions and prospects}
\label{conclu}

We have shown that the classical pure spinor string in the
$AdS_5\times S^5$ background has the full set of classical non-local
symmetries extending those recently studied by Klose, Loebbert and M\"unkler
in the context of ${\mathbb Z}_2$ cosets \cite{Klose:2016uur}.
We find that the inclusion of ghosts in a sense makes the ${\mathbb
  Z}_4$ symmetry manifest, and all non-local symmetries can be
lifted to the super-coset $PSU(2,2|4)/SO(1,4)\times SO(5)$.

An immediate extension of the results of this paper is to derive the
analog for the Green-Schwarz superstring. That can be done by
erasing the ghosts and imposing an appropriate gauge choice. Classical
solutions of the pure spinor string should preserve BRST symmetry
which means the BRST charge should vanish when evaluated on the
solution. If we do not set the ghosts to zero, this means that the
currents $K_3$ and $\bar K_1$ should vanish. In reference \cite{Vallilo:2003nx},
it was shown that the pure spinor flat current is equivalent to the
one in the Green-Schwarz formalism \cite{Bena:2003wd}\ in this gauge.
We expect that the GS string enjoys all of the symmetries discussed in the
present work.

It would be interesting to apply the results of this paper to
supersymmetric Wilson loops in $AdS_5$ as in
reference \cite{Klose:2016uur}. However, it is as yet not known how to study
%this class of
such classical solutions in the pure spinor formalism. There
is hope such a task can be done, since it was shown by explicit
computations in references \cite{Aisaka:2012ud,Cagnazzo:2012uq}
that the semi-classical quantization of the pure spinor string
is equivalent to the Green-Schwarz string in a certain class of
solutions. In \cite{Tonin:2013uec} it was argued that the equivalence holds
for any physical solution. With these results in mind, it is
likely that one can extend the results of, for example, references
\cite{Muller:2013rta,Munkler:2015gja} to the pure spinor string.

\vskip 0.3in
{\bf Acknowledgements.}
The work of O{\sc c} and B{\sc cv} is partially supported by FONDECYT
grant number 1151409. B{\sc cv} also has partial support from CONICYT grant
number DPI20140115. W{\sc dl}3 is supported by NSF grants
PHY-1214333 and PHY-1521099.
% We would like to thank Brodin for his blessing
% throughout all this work. We also thank Starbucks for hospitality.

\appendix
{\small
\bibliography{mybib}{}
\bibliographystyle{abe}
}

\end{document}